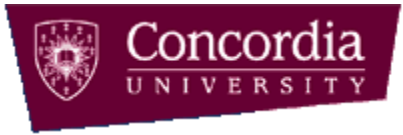

**COMP 6231: Distributed System Design**

**Technical Report**

# Fault-Tolerance through Message-logging and Check-pointing:

*Disaster Recovery for CORBA-based Distributed Bank Servers*

by
[1]Emil Vassev, [2]Que Thu Dung Nguyen and [3]Heng Kuang

[1]*emil@vassev.com*, [2]*nguyen@encs.concordia.ca*, [3]*h_kuang@encs.concordia.ca*

April 15, 2006

# Table of Contents

# Table of figures

# Team Members Contribution

| Name | Tasks |
|---|---|
| Emil Vassev | System architecture design.<br>BankServer design and implementation.<br>System integration. |
| Que Thu Dung Nguyen | System architecture design.<br>Recovery Module design and implementation. |
| Heng Kuang | System architecture design.<br>Monitor design and implementation. |

# 1. Introduction

## 1.1. Purpose of this document

In a distributed banking system, if the processor of a branch server crashes, access to information about the pending transactions in that branch is not only lost while the processor remains unavailable, but can also be lost forever. Message logging is one of the possible mechanisms that a branch server may use to recover information concerning that branch's accounts after a faulty processor has been repaired.

This document presents a failure-recovery mechanism for distributed bank servers.

## 1.2. Objective

Our goal is to build a failure-recovery variant of a Bank Server (BS) that provides fault tolerance features through message logging, and checkpoint logging and to build a middleware with Recovery Module (RM) and Monitor Module (Monitor) to support the recovery process of Bank Servers recovering from crash.

In this group of projects, three components are built to satisfy the requirements: 1) a message-logging protocol for the branch servers of the distributed banking system that logs the required information; 2) a Recovery Module that restarts the Bank Server using the log so that the restarted Bank Server can process subsequent requests for the various operations; and 3) a Monitor Module that provides support for the Bank Server via periodical checks whether the Bank Server is down and informs the Recovery Module to restart the Bank Server if the latter has crashed.

## 1.3. Scope

This system is built for the CORBA Distributed Bank Server (DBS). We assume that both Recovery Module and Monitor Module supporting the Bank Server are stable, i.e., we do not do recovery for these two modules.

The communication among all modules is based on the Transmission Control Protocol (TCP), which is a connection-oriented reliable protocol. This protocol detects errors, and we can easily detect if a message is not delivered to a recipient. Hence, we can build a reliable connection mechanism based on TCP.

The Distributed Bank Server system is built to be deployed on LAN.

# 2. Specifications

## 2.1. Message Logging

Message logging is a common technique used to build systems that can tolerate process crash failures. These protocols require that each process periodically records its local state and log the messages received since recording that state. When a process crashes, a new process is created in its place: the new process is given the appropriate recorded local state, and then it replays the logged messages in the order they were originally received.

All message-logging protocols require that the state of a recovered process be consistent with the states of the other processes. This consistency requirement is usually expressed in terms of *orphan processes* [1], which are surviving processes whose state is inconsistent with the recovered state of a crashed process. Therefore, the big question is how our implementation of message logging will guarantee that after recovery no process is orphan.

There are two possible solutions to that problem – a **careful logging** or a **complex recovery protocol**.

## 2.2. Bank Server

Bank Server handles in multitasking manner requests from Clients and other Bank Servers. The client's request could be open account, deposit, withdraw, balance and transfer. Other Bank Servers can request a participation in a transfer operation, which operation requires two bank servers. In addition, Bank Server implements fault tolerance ability through **careful message logging** and **checkpoint**. In case of a crash, the bank server has the ability to recover to the state as it is before crashing. The recovery process is part of the Bank Server's start process, which excludes handling of any out-coming request during the recovery. Bank Server communicates with the Monitor and Recovery Module for the purpose of recovery (for the design details of this module see section 4.1).

## 2.3. Recovery Module

Recovery Module always listens to requests from all Bank Servers and Monitor Module. Recovery Module processes the received requests in a multitasking manner. In addition, Recovery Module maintains the pool of dependency of Bank Servers, which dependencies are used for coordinated checkpoint of the dependent Bank Servers. Recovery Module forces all Bank Server in the pool of dependency to do checkpoints if a Bank Server requests to do a checkpoint. Recovery Module has the checkpoint operation synchronized among all the Bank Servers in a dependency. Recovery Module maintains the FIFO channel when accepting request doing checkpoint from different Bank Servers. Recovery restarts the Bank Server crashed as informed by Monitor Module (for the design details of this module see section 4.2).

## 2.4. Monitor Module

The main purpose of Monitor Module is to periodically check the Bank Server process and restart it if necessary. Monitor Module always listens to the heartbeat messages from all Bank Servers for a specified checking interval. Monitor Module registers a new Bank Server when receiving a register message. Monitor Module checks the heartbeat message time for each registered bank server. Monitor Module process these heartbeat messages as multitasking. Monitor Module communicates with Recovery Module and each Bank Server by corresponding thread with socket connection. Monitor Module sends a restart message to Recovery Module when certain bank server crashes (for the design details of this module see section 4.3).

# 3. System Architecture Design

## 3.1. Rationale

This is a distributed bank server application which is installed in a LAN. The application can support up to 100 Bank Servers. Bank Servers can run on separate hosts or can share the same host machines in the LAN. Each Bank Server binds its object reference in naming as "BankRemote" concatenating with its **branch number**. Clients use this name service to send requests to the desired Bank Server. Each Bank Server uses a port number as the same number as its branch number to listen to that port for requests from other Bank Servers that wish to do the transfer operation with it.

Recovery Module and Monitor Module can be installed on the same or separate host in the same LAN. Every Recovery Module uses two ports for communication with the Monitor and Bank Servers. Port 3000 is used to listen to general requests from other Bank Servers and the Monitor Module. Port 3001 is used for private communication between the Recovery Module and the Bank Servers for processing a checkpoint request by following the principles of 2-phase commit protocol.

Bank Servers send regular in time heartbeat messages to the Monitor for notifying they are active. The Monitor uses port 3100 to listen to those heartbeat messages coming from all the Bank Servers.

Bank Servers save to a persistent storage (file) their state and to a separate file their account operations as message logs. Therefore, each Bank Server maintains two files – one for message logs and one for the state (accounts and their balances).

Users view the provided interface and the result through the command line.

## 3.2. Work Flow

Recovery Module and Monitor Module are assumed to be stable and set up first. They always keep listening to all coming requests.

A Bank Server, when run, first finds the next available port within a range of 1000 (1111-2111) ports and starts with the branch number equal to this port. Next, the Bank Server starts the recovery process as follows:

- It loads its registered state by reading the Checkpoint file to obtain all the existing accounts associated with their amount if applicable.
- Next, it loads the registered messages from the message log file and re-executes them.

During the recovery process the Bank Server does not receive or process any incoming messages. But this messages if any are not lost they are simply accumulate in the port queue. After the recovery process is done the current state of the Bank Server is loaded in memory. Then the bank server spawns two threads:

- The first thread is for receiving requests from Bank Clients, other Bank Servers, and the Recovery Module.
- The second thread is for registering the Bank Server with the Monitor and for sending to the Monitor heartbeat messages every 30 seconds, this informing the Monitor for being the Bank Server alive.

Through the provided interface, client requests operation within its branch or involved with other branch (transfer). If operations are within the branch, Bank Server records all the detailed operation into the Message Logging file for recovery purpose. Meaning that these operations make changes to the state loaded from the Checkpoint Logging file. The current state of the branch is updated in the memory. If the operation is a transfer message logging is maintained by both participating Bank Servers. If the transfer operation succeeds, the leading Bank Server informs the Recovery Module for a dependency between the two Bank Servers participating in the transfer operation (for more details on the transfer operation see section 4.1.3).

If the Message Logging file maintained by a Bank Server comprises more 10 or more records, the Bank Server sends a requesting checkpoint message to the Recovery Module. Recovery Module retrieves the dependency vector of all Bank Server dependent on the requesting Bank Server and forces all the dependent Bank Servers to do checkpoint.

Monitor Module only needs to keep listening to heartbeat messages from all existing Bank Servers. If more than 30 seconds, Monitor does not receive any message from any of the Bank Server, it assumes that Bank Server crashed. Monitor informs Recovery Module the Bank Server crashed so the Recovery Module can restart that Bank Server.

Recovery Module receives dependent message, checkpoint message from Bank Servers and restart message from Monitor Module as mentioned above at port 3000. Recovery Module uses another port, port 3001, for 2-phase checkpoint commit with Bank Servers dependent on each other. After the 2-phase commit checkpoint succeed, all these Bank Server finished to do a new checkpoint, which recording their new state by writing down their state from memory to Checkpoint Logging file, delete the Message Logging file. And Recovery Module removes this dependency from the pool of dependencies it currently maintains.

## 3.3. Development Environment

The system was developed in Java using Eclipse software for programming. The JDK complier 5.0 is required.
The application can run on any Java 5.0 supporting environment.

## 3.4. Architecture Diagram

### 3.4.1. Component diagram
The application consists of three components.

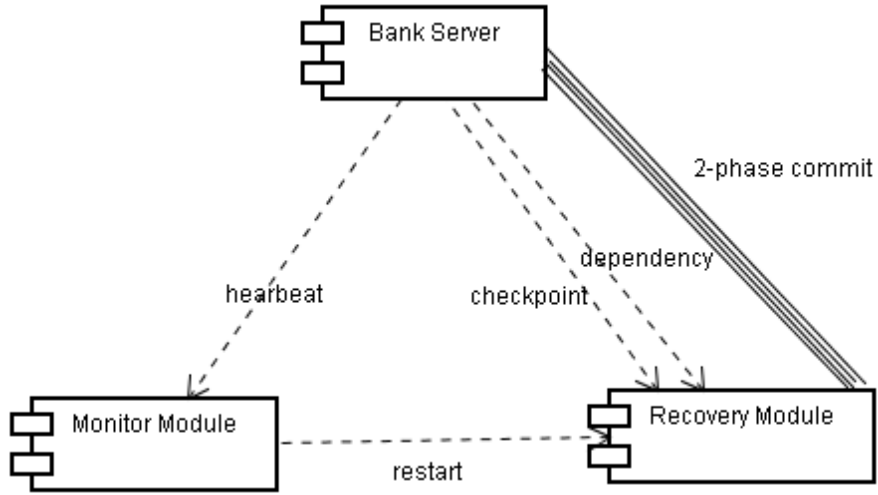

**Figure 1: System Component Diagram**

### 3.4.2. System layer diagram

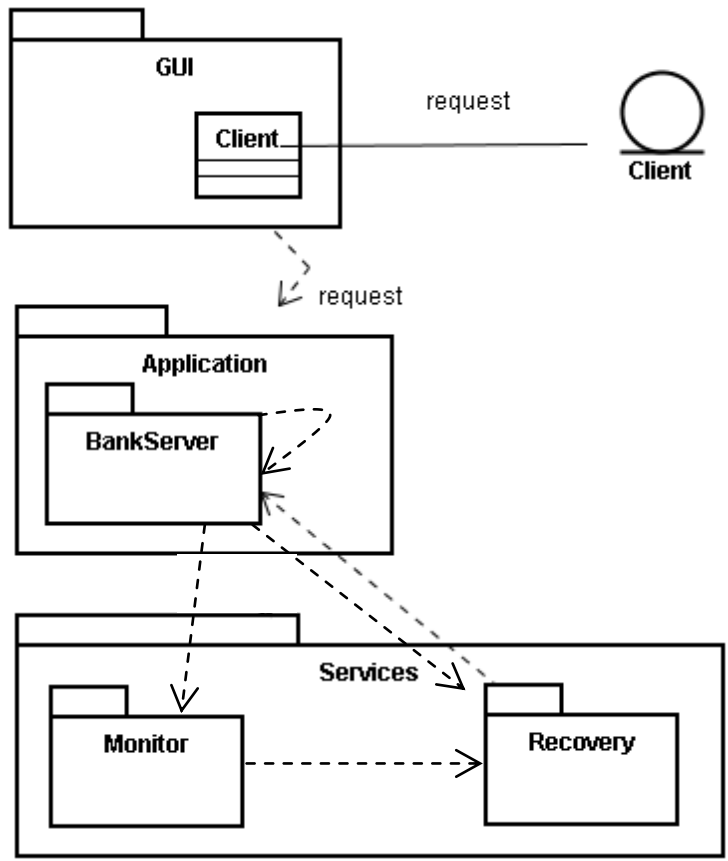

**Figure 2: System Layer Diagram**

# 4. Modular Design

## 4.1. Bank Server

### 4.1.1. The Recovery Process

In the course of this project we used message logging and checkpoint techniques to provide fault tolerance. This involves a recovery process, which process recovers the overall system and particularly the Bank Server (BS) that failed to a consistent state, recorded during the BS functioning. The following elements reveal the major details of the BS recovery process.

#### 4.1.1.1. Message-Logging and Checkpoints

In our design we use message logging and checkpoints to provide an effective recovery state mechanism. All the DBS inter-module communication goes through messages. Each message received by a BS is logged on a stable storage. Also, on a stable storage we log occasionally the BS current state, i.e. we log the BS accounts. The former operation refers to a **Message-Logging** operation and the last refers to a **Checkpoint** operation.

In the course of this project we defined a Message-Logging protocol and Checkpoint protocol, which protocols were needed for designing the bank server. These protocols define and formalize the structures of the **message log** and the **checkpoint log**.

**Message-Logging Protocol**
A message log records only messages with site effect, i.e. message that change the bank server state. Therefore the messages *"open account"*, *"deposit"*, *"withdraw"*, and *"transfer"* can be recorded in a message log. For recording these messages we specified the following format:
- All records begin with the tag BANK followed by the BS number and the colon sign ":".
- An *"open account"* message is recorded with the word OPEN followed by the account number.
  ***BANK #1111:OPEN 1111006***
- A *"withdraw"* message is recorded with the word WITHDRAW followed by the account number and the amount.
  ***BANK #1111:WITHDRAW 1111006 100.0***
- A *"deposit"* message is recorded with the word DEPOSIT followed by the account number and the amount.
  ***BANK #1111:DEPOSIT 1111006 1000.0***
- A "transfer" message is recorded with few records:
  - It begins with start transfer record, recorded with the words TRANSFER START followed by the source and destination accounts separated by the dash sign "-".
    ***BANK #1111:TRANSFER START 1111006-111200***

- Next follow the records describing the local account operations involved in the transfer operation. Those records could be WITHDRAW or DEPOSIT or both.
- It ends with commit transfer or cancel transfer record, recorded with the words TRANSFER COMMIT or TRANSFER CANCEL followed by the source and destination accounts separated by the dash sign "-".
  **BANK #1111:TRANSFER CANCEL 1111006-111200**
  or
  **BANK #1111:TRANSFER COMMIT 1111000-1112000**

The following is an example of a complete message log file:

**BANK #1111:OPEN 1111006**
**BANK #1111:DEPOSIT 1111006 1000.0**
**BANK #1111:WITHDRAW 1111006 100.0**
**BANK #1111:TRANSFER START 1111006-111200**
**BANK #1111:TRANSFER CANCEL 1111006-111200**
**BANK #1111:TRANSFER START 1111000-1112000**
**BANK #1111:WITHDRAW 1111000 10.0**
**BANK #1111:TRANSFER COMMIT 1111000-1112000**

**Checkpoint Protocol**
A checkpoint log records only accounts' state for the checkpointed bank server (BS). For recording the accounts' state we specified the following format:
- All records begin with the tag BANK followed by the BS number and the colon sign ":".
- The account information is stored as account number followed by the current account's balance.
  **BANK #1111:1111005 1030.0**

- Each account is recorded on a separate line.

The following is an example of a complete checkpoint log file:

**BANK #1111:1111005 1030.0**
**BANK #1111:1111004 0.0**
**BANK #1111:1111003 1030.0**
**BANK #1111:1111002 120.0**
**BANK #1111:1111001 130.0**
**BANK #1111:1111000 1374.0**

## 4.1.1.2. Recovery State Algorithm

As the system runs, new logged messages and checkpoints arrive on the stable storage. <u>A combination of checkpoint log and message log defines the recovery state for every bank server.</u>
In our design the execution of bank servers between received messages is deterministic. This helped us to design a recovery process based on message logging and checkpoints. The recovery process includes the following sequence of steps performable on the bank server side:

- Bank servers log any message with site effect (message that change the bank server state – like deposit, withdraw and transfer). Logging is storing messages into a stable storage.
- Logging should be done right after processing the message, this allowing keeping consistent the message log with the performed operations.
- Bank servers regularly notify the Monitor for being active.
- Bank servers are occasionally check-pointed to a stable storage (accounts are stored on the hard disk). The checkpoints are requested by the bank servers when the size of their message log reaches a predefined critical point (for example the message log file should not store more than 50 log records).
- If a checkpoint is performed by a bank server, the last deletes its message log file and starts new one immediately after finishing with the checkpoint operation.
- If a bank server failed:
  - The Monitor stops receiving notifying messages from this server and request the RM to restart the server.
  - The bank server restarts from its last checkpoint.
  - All the messages, received by this bank server since the last checkpoint, are read from the message log file and replayed in the same order.
  - The bank server re-executes based on these messages to its last consistent state – just before the time of failure.

<u>The recovery process takes place in the Bank Server constructor. At his time the Bank Server (BS) is not allowed to receive and process new messages. This ensures a consistent recovery process, since there is no possibility for BS state changes except those directed by the recovery process.</u>

## 4.1.1.3. Garbage Collection

When a bank server (BS) works, it stores checkpoint and logged messages into files on the hard disk - stable storage in case they are needed for some future recovery. The stored data may be removed from the stable storage, whenever doing so will not interfere with the ability of the system to recover as needed.
In our bank server design:

- the **checkpoint log file** must be created after the first checkpoint operation performed by a bank server and this checkpoint log file must be refreshed with the new BS state in any consecutive checkpoint operation but never deleted;

▪ the **message log file** must be deleted after each checkpoint done by the BS.

## 4.1.2. Message Exchange with Other DBS Modules

A bank server exchanges multiple messages with the other DBS modules – bank servers (BS), recovery module (RM) and Monitor. All the messages are exchanged via the Transmission Control Protocol (TCP), which is a connection-oriented **reliable** protocol. This protocol detects errors. Therefore, we can easily detect message delivery failures. The presence of many messages exchangeable between BS and the other bank system applications requires careful message-passing design and implementation. The following elements depict the message structure of the messages designed and implemented in the course of this project.

### 4.1.2.1. BS-BS Messages

**Message TRANSFER**
**Usage:** This message takes place in Transfer operations. It is sent by the leading bank server (the one contacted by the client) to the second bank server.
**Format:** {BS#} TRANSFER {Source Account #} {Destination Account #} {Amount}

**Message OK**
**Usage:** This message takes place in Transfer operations. It is sent by the second server back to the leading server in response to the TRANSFER message.
**Format:** {BS#} OK

**Message ERROR DESCRIPTION**
**Usage:** This message takes place in Transfer operations. It is sent by the second server back to the leading server in response to the TRANSFER message. The message describes the exception raised (if any) during the local account operation.
**Format:** {BS#} {Exception}

### 4.1.2.2. BS-RM Messages

Considering the design of message passing between BS and RM we decided to use two ports for BE-RM messaging – 3000 and 3001. All the messages concerning the checkpoint operations are going via socket port 3001. The rest are going via socket port 3000. This helped us to keep RM receiving standard messages, while working on a checkpoint.

**Message DEPENDENCY**
**Usage:** This message is sent right after a successful *Transfer* operation. The leading server sends the DEPENDENCY message to the RM. The message notifies the RM for the dependency between two servers – BS1 and BS2, which dependency is a consequence of the *Transfer* operation between those servers. Dependencies help with the checkpoint operations, where in order to preserve the overal system consistency all the dependent bank servers must do a check point at the same time. This message goes via port 3000.

**Format:**  DEPENDENCY  {BS1#}  {BS2#}

**Message CHECKPOINT**
**Usage:** This message is sent by the bank server to the RM. The message is a request for a checkpoint, i.e. the sender requests a checkpoint that must be maintained by the RM, since all the dependent bank servers should do a check point as well. This message goes via port 3000.
**Format:**  CHECKPOINT  {BS#}

**Message READY_FOR_CHECKPOINT**
**Usage:** This message is a "ready for checkpoint" request sent by the RM to the bank servers participating in a currenlty going checkpoint. Also, the bank servers send this message back to the RM as a response to the "ready for checkpoint" request.  This message goes via port 3001.
**Format:**  READY_FOR_CHECKPOINT

**Message DO_CHECKPOINT**
**Usage:** This message is sent by the RM to the BSs that confirmed ready for checkpoint. This message goes via port 3001.
**Format:**  DO_CHECKPOINT

**Message CANCEL_CHECKPOINT**
**Usage:** This message is sent by the RM to the BSs to cancel the checkpoint operation. This message goes via port 3001.
**Format:**  CANCEL_CHECKPOINT

**Message CHECKPOINT_DONE**
**Usage:** This message is sent by the BSs to the RM to confirm they are done with the checkpoint operation. This message goes via port 3001.
**Format:**  CHECKPOINT_DONE

### 4.1.2.3. BS-Monitor Messages

**Message REGISTER_MSG**
**Usage:** This message takes place in the communication between the bank servers and Monitor. It is sent by each bank server at its starting time in order to register with the Monitor as a sender of heartbeat messages.
**Format:**  REGISTER_MSG  {BS IP address}  {BS#}

**Message HEARTBEAT_MSG**

**Usage:** This message takes place in the communication between the bank servers and Monitor. It is sent by each bank server in a regular fashion to the Monitor. The message notifies the Monitor that the sender is active bank server and does not need to be restarted. If a BS fails to send this message for a specified amount of time, the Monitor will request the RM to restart that BS.

**Format:**  HEARTBEAT_MSG  {BS#}

## 4.1.3. Transfer Operation with Careful Logging and Built In Rollback

The Transfer operation is more complex and it requires **careful logging**. If the sender (bank server #1) fails and cannot be fully recovered (for example because the transfer message has not been logged) the receiver (bank server #2) becomes an *orphan*, and its state must be rolled back during recovery to a point before this dependency was created. If rolling back causes other bank servers to become *orphans*, they too must be rolled back during recovery – domino effect. This is the worst case scenario, which scenario we want to avoid. In order to avoid *orphans* among the bank servers we implement the *Transfer* operation with **careful logging**. This allows servers doing a Transfer operation to maintain a **build in roll back** in case the transfer operation fails. Hence, we do avoid *orphans* because the Transfer operation is always consistent for both sides (both servers participating in this Transfer operation).

If a *transfer* operation occurs between two bank servers – BS#1 and BS#2, the operation is registered on both message logs as a sequence of records:
  1. Start Transfer Record
  2. Operation Records
  3. Commit/Cancel Transfer Record.

The operation records are *withdraw* and *deposit* operations. For a successful *transfer* operation we register *withdraw* operation for BS#1 and a *deposit* operation for BS#2. The following samples show a transfer operation registered on both sides:

**Message Log for a Successful Transfer Operation – BS#1**
*BANK #1111:TRANSFER START 1111000-1112000*
*BANK #1111:WITHDRAW 1111000 10.0*
*BANK #1111:TRANSFER COMMIT 1111000-1112000*

**Message Log for a Successful Transfer Operation – BS#2**
*BANK #1112:TRANSFER START 1111000-1112000*
*BANK #1112:DEPOSIT 1112000 10.0*
*BANK #1112:TRANSFER COMMIT 1111000-1112000*

If BS#2 fails during the Transfer operation, BS#1 will do a rollback, which is a *deposit* operation following right after the *withdraw* operation:

**Message Log for a Failed Transfer Operation – BS#1**
*BANK #1112:TRANSFER START 1112000-1111000*

*BANK #1112:WITHDRAW 1112000 10.0*
*BANK #1112:DEPOSIT 1112000 10.0*
*BANK #1112:TRANSFER CANCEL 1112000-1111000*

Since, the server that crashed did not complete the transfer operation the message logging for this operation for that server is not complete as well. The message logging below does not include a *Commit/Cancel* record:

**Message Log for a Failed Transfer Operation – BS#2 (Crashed Server)**
*BANK #1111:TRANSFER START 1112000-1111000*
*BANK #1111:DEPOSIT 1111000 10.0*

Hence, during the recovery process BS#2 will not find the enclosing transfer operation record – *Transfer Commit/Cancel*, and will simply not perform the account operation (deposit in this case) and this is just fine, since, BS#1 has already done a rollback.

The following elements depict a general scenario for a transfer operation with **careful logging** and built in rollback. Both bank servers – BS#1 and BS#2, that participate in the transfer operation maintain their own message log file – *msglog1* and *msglog2* respectively:

1. A bank clIent sends a transfer request to BS#1.
2. BS#1 writes a "Transfer Start message" into the message log file *msglog1*.
3. BS#1 does a withdraw operation.
4. BS#1 writes a "Withdraw message" into the message log file *msglog1*.
5. BS#1 sends a message "Transfer" to BS#2.
6.1. BS#1 starts waiting for the response of BS#2.
   - If the response is OK, BS1 does a commit operation:
     a. BS#1 writes a "Commit" message into the message log file *msglog1*.
     b. BS#1 sends a dependency message to RM.
   - If the response is an error, BS#1 does a rollback operation:
     a. BS#1 does a deposit operation.
     b. BS#1 writes a "Deposit message" into the message log file *msglog1*.
     c. BS#1 writes a "Cancel message" into the message log file *msglog1*.
   - If there is no response due to BS#2 crash or network failure, BS#1 does a rollback operation:
     a. BS#1 does a deposit operation.
     b. BS#1 writes a "Deposit message" into the message log file *msglog1*.
     c. BS#1 writes a "Cancel message" into the message log file *msglog1*.
6.2. BS#2 writes a "Transfer Start message" into the message log file *msglog2*.
7. BS#2 does a deposit operation.
8. BS#2 writes a "Deposit message" into the message log file *msglog2*
9. BS#2 sends an "OK message" to BS#1.
   - If the message is delivered to BS#1, BS#2 does a commit operation:
     a. BS2 writes a "Commit" message into the message log file *msglog2*.
   - If the message cannot be delivered to BS#1, BS#2 does a rollback operation:
     a. BS2 does a withdraw operation.
     b. BS2 writes a "Withdraw message" into the message log file *msglog2*.

    c. BS2 writes a "Cancel message" into the message log file *msglog2*.

The following elements depict a base success scenario for a transfer operation with **careful logging** (see Fig. 3). Both bank servers – BS#1 and BS#2, that participate in the transfer operation maintain their own message log file – *msglog1* and *msglog2* respectively:

1. A bank cleint sends a transfer request to BS#1.
2. BS#1 writes a "Transfer Start message" into the message log file *msglog1*.
3. BS#1 does a *withdraw* operation.
4. BS#1 writes a "Withdraw message" into the message log file *msglog1*.
5. BS#1 sends a message "TRANSFER" to BS#2.
6. BS#2 writes a "Transfer Start message" into the message log file *msglog2*.
7. BS#2 does a *deposit* operation.
8. BS#2 writes a "Deposit message" into the message log file *msglog2*.
9. BS#2 writes a "Transfer Commit message" into the message log file *msglog2*.
10. BS#2 sends a message "OK" to BS#1.
11. BS#1 writes a "Transfer Commit message" into the message log file *msglog1*.
12. BS#1 sends a message "DEPENDENCY" to RM.

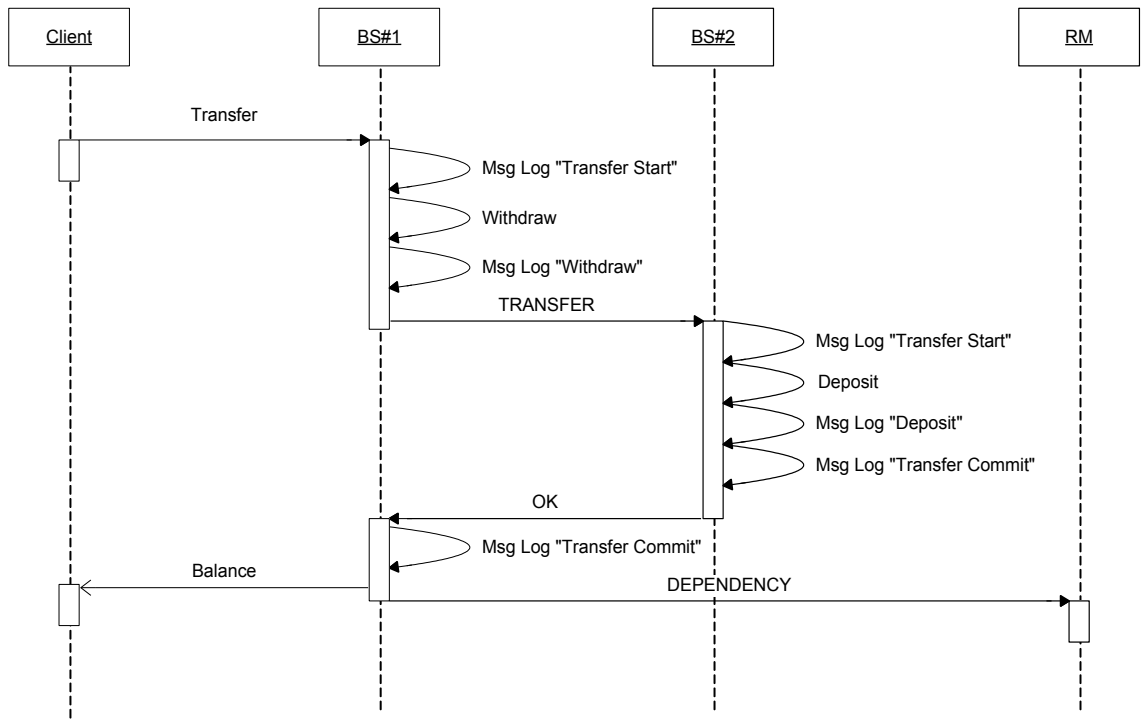

**Figure 3: Transfer Base Success Scenario – Sequence Diagram**

### 4.1.3. Design

The bank server application creates *remote bank objects* (instances from the *BankRemoteImpl* class) and exposes these objects to the clients via a remote interface called *BankRemote.IDL* and its Java's analog *BankRemote* interface (see Fig. 4). Here, the interface defines methods, and the class implements those methods.

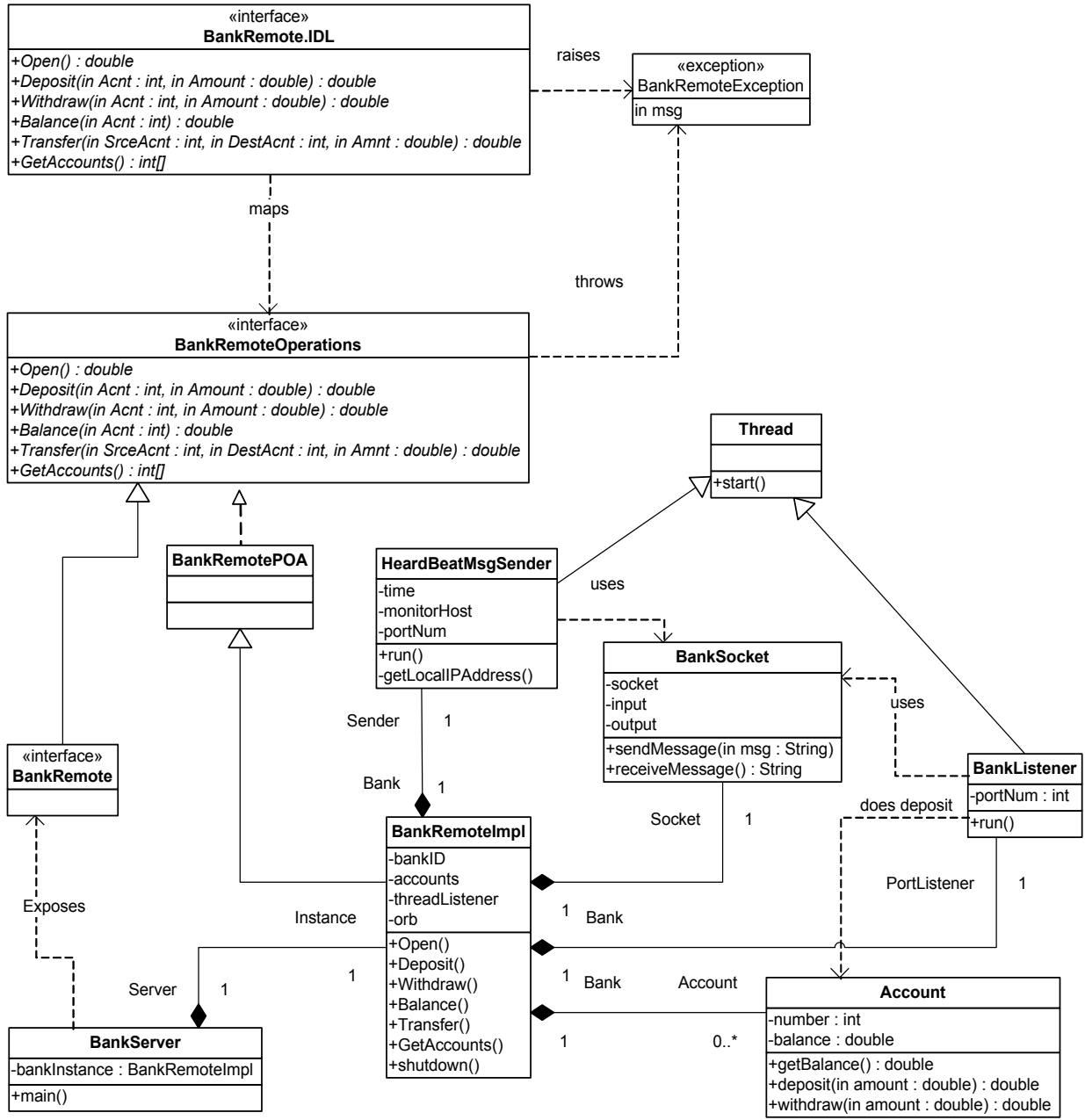

**Figure 4: Bank Server Class Diagram**

The bank server program consists of two major classes - the *servant* (see *BankRemoteImpl* class) and the *server (*see *BankServer* class). The servant,

*BankRemoteImpl*, is the implementation of the *BankRemote IDL* interface, i.e. each *BankRemote* instance is implemented by a *BankRemoteImpl* instance. The *servant* is a subclass of *BankRemotePOA*, which is generated by the **idlj** compiler from the *BankRemote.IDL*. The *servant* contains one method for each IDL operation (see Fig. 4) – *Open(), Deposit(), Withdraw(), Balance(), Transfer()* and *GetAccounts()* methods. *Servant* methods are ordinary Java methods with some extra code to deal with CORBA user defined exceptions. The functionality needed for dealing with ORB, marshaling arguments and results, etc. is provided by the skeleton.

### 4.1.3.1. Design Rationale

The Design Rationale (DR) for the CORBA bank application expresses elements of the reasoning, which has been invested during the design process. A DR answers "*Why...?*" questions of different sorts, depending on the class of DR represented.

CORBA ensures that two processes running on separate machines can exchange invocation requests and results [2]. The use of CORBA requires a design of a special IDL interface (see BankRemote.IDL) using the OMG's Interface Definition Language (IDL). IDL has a syntax similar to C++ and can be used to define modules, interfaces, data structures, and more. The IDL can be mapped to a variety of programming languages including Java, this ensuring **interoperability** in CORBA.

Hence, by defining the CORBA IDL interface and programming against this interface, in our solution we have applied static invocation, which uses a client stub for the invocation and a server skeleton for the service being invoked.

Whereas the implementation of the operations *Open, Deposit, Withdraw, Balance,* and *Get Accounts* does not require other considerations than the standard ones for concurrent programming with Java and CORBA, the implementation of the *Transfer* operation requires TCP socket programming as well.

As an answer to the requirements of the fail-recovery CORBA bank server we propose an IDL interface and six classes (See Fig. 4):
- IDL interface *BankRemote* – defines the access to the CORBA *remote bank object*.
- Class *BankServer* – implements the bank server application that creates *remote bank objects* and exposes them to the clients via a remote interface.
- Class *BankRemoteImpl* – defines a CORBA *remote bank object* by implementing the methods of the CORBA IDL interface *BankRemote*.
- Class *BankSocket* – wraps the Java socket class for TCP connection between two bank servers.
- Class *BankListener* - instantiates threads within a CORBA *remote bank object*. This threads establish a socket connection with another CORBA *remote bank object*.
- Class *HeardBeatMsgSender* - instantiates threads that send heartbeat messages to the Monitor.
- Class *Account* – encapsulates the functionality for an account.

### 4.1.3.2. Concurrency Issues

Our design takes into account possible concurrent execution of many instances of the same bank server, those working on the same set of accounts and the same message log and checkpoint files. This is coming from the fact that CORBA spawns many bank server threads from the same BS, i.e. one BS thread per client.
Fig. 5 depicts the concurrent execution of two thread instances of bank server BS1.

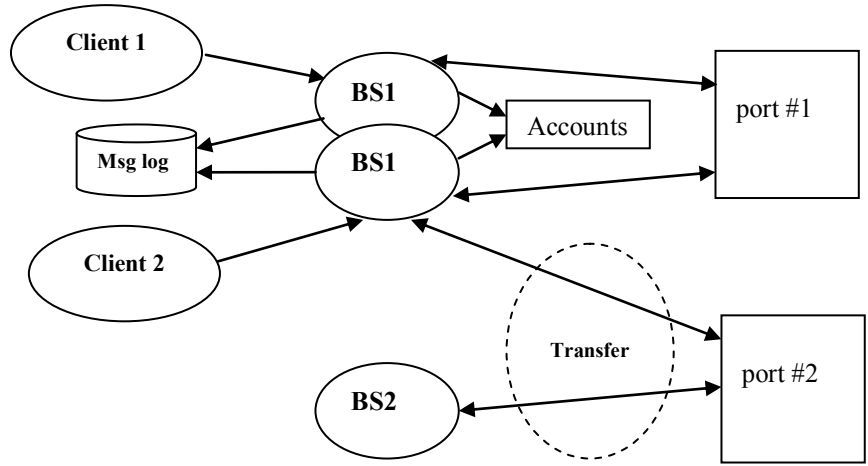

**Figure 5: Concurrent Execution of Two BS Threads**

The two BS1 threads share accounts, ports, message log files etc. (see Fig. 5). Considering that, we designed our Bank Server with all the account-related and exclusive-access port operations synchronized. For example, we cannot perform simultaneously two Transfer operations from the same BS. Also, the operations for writing in the message log file are synchronized as well. Therefore, if there is currently going writing all the BS1 threads will be synchronized on it.

### 4.1.3.3. Interface BankRemote.IDL

This interface is written in CORBA IDL language. An interface definition written in CORBA IDL completely defines the interface and fully specifies each operation's parameters by ensuring platform **interoperability**. Therefore, this interface specifies the methods that can be invoked *remotely* by a bank client. These methods are:
- *Open()* – this method opens a new account and returns the account number.
- *Deposit()* – this method causes the balance of an account to be increased by a specified amount and returns the account's balance.
- *Withdraw()* – this method causes the balance of an account to be decreased by a specified amount and returns the account's balance.
- *Balance()* – this method returns the current balance of a specified account.
- *Transfer()* – this method causes the balance of an account to be decreased by specified ammount and the balance of another account to be increased by the same amount.

- *GetAccounts()* – this method returns an array of account numbers for all the accounts in the bank.

The methods *Deposit(), Withdraw(), Balance(),* and *Transfer()* throw an exception *BankRemoteException*, defined within the IDL interface:

```
exception BankRemoteException {
    string msg;
};
```

This exceprion is mapped to an extension of the org.omg.CORBA.UserException exception and it is triggered by the *remote bank objects* (*RemoteBankImpl* class instances). This exceptions contains user-defined data – the *msg* attribute, <u>which is necessary to read the message (the reason caused that exception)</u>. When handling this exception, getting the error message with the standard Java exception's method *getMessage()* does not return the appropriate message. Hence, we use:

```
catch (Bank.BankRemotePackage.BankRemoteException ex)
{
    System.err.println("Bank Server error: " + ex.msg);
    pressEnter();
}
```

The *BankRemoteException* exception implementation is generated uatomatically by the **idlj** compiler.

In order to define with IDL the *GetAccounts()* method result type, which is an array of integers, we defined an *IDL sequence* of type long. An *IDL sequence* is similar to a one-dimensional array of elements, but it does not have a fixed length.

```
interface BankRemote
{
    typedef sequence<long> Accounts;
    …
    Accounts GetAccounts();
}
```

The IDL interface was mapped to a java interface with the **idlj** compiler, which mapping generated a bunch of files including: _BankRemoteStub.java, BankRemote.java, BankRemoteHelper.java, BankRemoteHolder.java, BankRemoteOperations.java, BankRemotePOA.java, BankRemoteException.java etc.

Clients program to remote interfaces, not to the implementation classes of those interfaces. Part of the design of such interfaces is the determination of any objects that will be used as parameters and return values for these methods.

### 4.1.3.4. Class BankRemoteIml

The class *BankRemoteImpl* extends the *BankRemotePOA* class, which is generated automatically from **idlj** compiler. *BankRemotePOA* is an abstract class referring to but not implementing the *BankRemoteOperations* interface, which is the Java analog of the *BankRemote IDL* interface (see Fig. 4). Hence, in our *BankRemoteImpl* we have to implement the methods exposed by *BankRemoteOperations* interface (respectively those exposed by the *BankRemote IDL* interface). Therefore, in BankRemoteImpl class we implement the methods *Open(), Deposit(), Withdraw(), Balance(), Transfer()* and *GetAccounts()*.

In addition, *BankRemoteImpl* class extends (via the *BankRemotePOA* class) the class *org.omg.PortableServer.Servant*, this allowing creation and export of remote objects via ORB. Exporting a remote object makes that object available to accept incoming calls from clients. ORB passes a remote *stub* for this remote object to the clients, this preventing the implementation of this object in the client's machine. The *stub* acts as the local representative (or proxy) for the remote object. The clients use that proxy locally and transparently as they use the real *BankRemoteImpl* object, i.e. they make method calls on the proxy. The proxy just transmits calls across the network to the real *BankRemoteImpl* object.

The class *BankRemoteImpl* class defines internally the classes *BankSocket, BankListener, HeardBeatMsgSender,* and *Account* (see classes *BankSocket, BankListener, HeardBeatMsgSender, and Account* below) as inner classes. Defining the last as inner classes hides these classes from a possible external use and allows the outer and inner classes to share their non-public declarations.
For example, for accessing the account's id, within the class *BankRemoteImpl* but outside the class *Account*, we simply write:

```
oAccount.iNum;
```

Therefore, with this design aspect we improve the data hiding, this improving the data consistency, but also we improve the interoperability between the classes *BankRemoteImpl* and *Account*. This is very important aspect, considering the extremely tight relationship between these two classes - the accounts belong to the bank, and cannot be shared among bank instances. Hence, the composition relationship between these two classes (see Fig. 4).

### The Constructor(s)
The constructors first call the super class constructor. The class *BankRemoteImpl* defines constructors accepting as arguments the Monitor's host address and eventually the Bank Server number (BS number). If the BS number is not provided the BS finds the first available port within the range {1111, 2111} and use the number of the found free port as a BS number. Next, the BS recovers its state:
- loads the last recorded BS state from the checkpoint log file;

- uses the message log file to load and re-execute side effect messages (if any) that have been processed after the last checkpoint.

```
//**** Loads the last checkpointed state
LoadState();

//**** Does Recovery from the message log if any.
bNotRecovery = false;
Recovery();
bNotRecovery = true;
```

Next the constructors spawn a *BankListener* thread for accepting incoming messages and a *HeardBeatMsgSender* thread for sending heartbeat messages (see *BankListener* class and *HeardBeatMsgSender* class below). Spawning those two threads after the recovery process makes the recovery process safe, since no newly coming messages can be processed during the recovery time.

```
//**** Spawns a BankListener thread.
//**** The port number is set equal to the bank id (sBranchNum)
oBankListener = new BankListener(iSocketPort);
oBankListener.start();

//**** Spawns a HeardBeatMsgSender thread.
oHeardBeatMsgSender = new HeardBeatMsgSender(MONITOR_PORT_NUMBER,
                                      30, psMonitorHost);
oHeardBeatMsgSender.start();
```

Also, spawning a separate thread for listening to the incoming socket port and a separate thread for sending heartbeat messages prevents the main thread from freezing on those two operations. Hence, our Bank Server is fully active – we can do all the operations on it, while it is listening to the incoming port for incoming messages, and while heartbeat messages are sending to the Monitor.

**The Message Logging**

The message logging is built in all the operations with site effect – open account, deposit, withdraw, and transfer. The functions performing those operations call the synchronized *WriteMessageLog()* function. This function stores the messages into the message log file. The operation is synchronized, this preventing concurrent message logging. A concurrent message logging could break the message consistency.

The *WriteMessageLog()* function increment a message counter after each message storing. If the counter reached a special threshold (set to 10 messages) a "CHECKPOINT" message will be sent to the RM.

```
private synchronized void WriteMessageLog(String psMsg)
{
…

//**** requests a checkpoint if there is no currently going transfer operations
      if ((!bInTransfer) && (iMsgCounter >= CHECKPOINT_THRESHOLD))
      {
            String sMsg = "CHECKPOINT " + sBranchNum;
            SendMessage(sMsg, sRMHost, RM_PORT_NUMBER);
…
```

**Checkpoint**

A checkpoint operation could be triggered by the Bank Server or by another Bank Server if there is a dependency among the bank servers. The checkpoint is performed by two methods *ReadyForCheckPoint()* and *doCheckPoint()* and the request for a checkpoint is performed by the *WriteMessageLog()* function (see the section above).

*ReadyForCheckPoint()* method covers the first part of the *Two-phase Checkpoint Protocol* (see section 4.2 Recovery Module). This method is called by the *run()* method of the *BankListener* class when a "READY_FOR_CHECKPOINT" message is received. <u>Since, the method is called from the thread listening for incoming messages, this thread is blocked until the end of the *ReadyForCheckPoint()* method. In addition this method is implemented as *synchronized*, which blocks all the bank account operations implemented as *synchronized* methods as well. Therefore, during the execution of this method the Bank Server is blocked except the thread sending regular heartbeat messages to the Monitor. In addition, this method calls the *doCheckPoint()* method which saves the BS state into a file. The *doCheckPoint()* method is *synchronized* as well. Hence, the checkpoint operation is safe.</u>

Both methods *ReadyForCheckPoint()* and *doCheckPoint()* communiocate with the RM via RM's second port 3001 (see section 4.2. RM).

Also, the method *ReadyForCheckPoint()* calls the synchronized method *ReadMessages()* for reading only checkpoint-related messages coming from the RM during the checkpoint operation. This method spawns intrenally a timeout featured thread for reading the expected message, i.e. if the message is not received for a specified time, the operation could be canceled due to timeout error.

**Recovery**

The recovery process is performed by two methods – *LoadState()* and *Recovery()*. The former simply loads the checkpoint log file and re-instantiate the bank accounts with their balances. The last loads and re-execute the messages stored in the message log file. These two methods are called by the constructors only – at the Bank Server construction time.

**Accounts**

For maintaining a pool of open accounts, the class *BankRemoteImpl* defines a hash table:

```
private Hashtable<Integer, Object> accounts = new Hashtable<Integer, Object>();
```

This hash table maps keys to values, were the keys are account numbers and values are *Account* objects corresponding to these numbers. This design aspect helps with maintaining a dynamic set of open accounts, i.e. an extensible set of accounts. Every client can create a new account by calling the *Open() BankRemote's* method, this adding the newly created (open) account to the hash table and making this account accessible by the other clients.

The class *BankRemoteImpl* implements the IDL interface *BankRemote*. Therefore, the former implements all the interface's methods *Open(), Deposit(), Withdraw(), Balance(), Transfer()* and *GetAccounts()*. The methods *Deposit(), Withdraw(), Balance()* and *Transfer()* throw *BankRemoteException* (see interface *BankRemote.IDL*) in cases there are some erroneous conditions like "account does not exist".

```
if (oAccount != null)
{
     synchronized (oAccount)
     {
          return oAccount.getBalance();
     }
}
else
     throw new Bank.BankRemotePackage.BankRemoteException ("Account
#" + piAcnt + " does not exists.");
```

These methods lock the *Account* object they operate on, by using synchronized block of execution. This ensures that a second call on these methods will be blocked until the first does not release the *Account* object (see class *Account*).

The *Transfer()* method is the most significant from implementation perspective:

- First, the method searches for the source account the pool of accounts maintained by the local server (see hashtable *accounts*). If the account is not found *BankRemoteException* is thrown.
- Second, the method checks if the funds of the source account are sufficient. If the funds are not sufficient *BankRemoteException* is thrown.
- Third, the method searches for the destination account the pool of accounts maintained by the local server (see hashtable *accounts*). If the account is found local transfer operation is performed, otherwise remote transfer operation is performed.
- Fourth, the method locks the source account. If the destination account is local it is locked as well – this preventing any out coming operations on those two accounts.
- Fift, performs a local or remote transfer depending on the destionation account location.

The local transfer is sequence of withdraw, deposit, and get balance operations performed on the locked local *Account* objects.
The remote transfer is more complex. It starts with a socket connection to the remote bank server. Next, we do kind of data wrapping – we put all the necessary info for a remote transfer in one message – a byte stream where the elements are separated by empty space:

```
sMessage = sBranchNum + " transfer " + piDestAcnt + " " + pdAmt;
```

Hence, when this message is received by the remote bank server, the last will find out that this is a "transfer" of amount *pdAmt* to account *#piDestAccount* ordered by bank *# sBranchNum*.
Next, we send the message via the socket connection and start waiting for the upcoming message. When this message is received we check if it si "OK" message. If it is not "OK" it is the remote error description, which we throw as a *BankRemoteException* exception.
For message logging of the Transfer operation see section 4.1.3.

### 4.1.3.5. Class BankSocket

This class wraps the Java Socket class, by providing *sendMessage* and *receiveMessage* functions. The class internally maintains the input and output buffers and takes precautions for buffer flushing. This class is an inner class for the *BankRemoteImpl* class. This desing aspect keeps the BankSocket class for internal use only, this preventing socket ports sharing among different *BankRemoteImpl* instances. Hence, the major issue here is improved object encapsulation, which is one of the most important aspects in CORBA programming.

### 4.1.3.6. Class BankListener

This class instantiates threads used for establishing socket connection with other Bank Servers (BS). Multithreading allows cocnurrent execution of the bank operations from one side and receiving messages from another side. The significant part of this class is the *run()* method implementation. This method:

- First, instantiates a socket for accepting connection.
- Second, it runs a loop for reading incoming messages.

```
while (!bEndThread)
{
…
}
```

This loop ends when the ORB is shut down (see the *shutdow*n() method of the *BankRemoteImpl* class). The *shutdow*n() method of the *BankRemoteImpl* class makes the *bEndThread* variable *true* and joins the *BankListener* thread before shutting down the server.

Within the loop, the *run()* method:

- First, waits to accept a connection request, at which time a data socket is created.
- Second, receives the incomming message.
- Third, parses the message.
- Fourth, if the message is a "TRANSFER" message:
  - BS does a deposit operation onto specified by the message account with the specified by the message amount.
  - If everithing is OK, creates an "OK" message and sends it back via the socket connection. If there is an error wraps this error in a message and sends it back via the socket connection.
- Fifth, if the message is a "READY_FOR_CHECKPOINT" message, BS process this messages by calling a special method.

### 4.1.3.7. Class **HeardBeatMsgSender**

This class instantiates threads that send heartbeat messages to the Monitor. By implementing this class as a thread, we address a multithreading issue that allows concurrent execution of the bank operations, listening to incomming messages and sending heartbeat messages in a regual time fashion. This class is aware about the Monitor's host address and incomming port. <u>In addition, the class maintains a time variable for the period of time used to measure the time elapsed between two heartbeat messages to be sent. This time is currently set to 30 sec.</u> The whole information needed by the class *HeardBeatMsgSender* is passed to the class via its constructor's arguments:

```java
public HeardBeatMsgSender(int piPortNum, int piTime, String psMonitorHost)
{
    iPortNum = piPortNum;
    iTime = piTime;
    sMonitorHost = psMonitorHost;
}
```

The *run()* method simply runs an infinite loop where the current time is measured in miliseconds and if the elapsed time since the last heartbeat message reached the time period (30 sec.) a new heartbeat message is sent to the Monitor by using an instance of the *BankSocket* class. The loop ends when the ORB is shut down (see the *shutdow*n() method of the *BankRemoteImpl* class).

### 4.1.3.8. Class **Account**

This class encapsulates the functionality of the bank account. This class defines methods *getBalance(), deposit(),* and *withdraw()*, which are declared as *synchronized* methods. In Java programming, each object has a lock acquiring with the *synchronized* keyword. Hence, the *Account*'s synchronized methods can only be executed by one thread at a time for a given instantiation of a class, because that code requires obtaining the object's lock before execution. In that sequence, the *BankRemoteImpl* object acquires the lock for *Account* objects by using the synchronized keyword in the following way:

```java
if (oAccount != null)
{
    synchronized (oAccount)
    {
        return oAccount.deposit(pdAmnt);
    }
}
else
    throw new Exception("Account #" + piAcnt + " does not exists.");
```

This desing aspect prevents the account's balance from inconsistency, when two or more clients work on the same account simultaneously. Therefore, this mechanism allows maintaining data consistency by ensuring the *oderly execution* of the account's methods by of cooperating clients.

This class maintains only two data fields – account number and account balance. For more complete information, other fields can be added like owner's id, date of creation, currency, etc.

The class *Account* is declared as inner class for the class *BankRemote*, this improving the overall data consistency (see class *BankRemoteImpl*).

### 4.1.3.9. Class BankServer

This class defines the bank server application. The `BankServer` class (see Fig. 4) defines the *main()* method, which acts as an entry point to the program. This method does the following:

- Creates and initializes an ORB instance
- Gets a reference to the root POA and activates the `POAManager`
- Creates servant and register it with the ORB.
- Creates a tie with the servant being the delegate
- Gets a CORBA object reference for a naming context in which to register the tie. This step also implicitly activates the object.
- Gets the root naming context .
- Registers the new object in the naming context under the name "BankRemote"
- Waits for invocations of the new object from the client.

## 4.2. Recovery Module

### 4.2.1. Composition

The design of Recovery Module is simple. It has two ports and maintains the pool of dependency.

Recovery Module has two ports. One port always listens to all requests coming from Bank Server and Monitor Module. Each time the first port receives a request, Recovery Module exams the request. If Recovery Module needs to communicate with a specific Bank Server to handle this request, it can do so with the second port. The first port becomes free, comes back and continues to listen to other request.

Recovery Module maintains a pool of dependency as another feature of it. When a Bank Server requests to do a checkpoint, Recovery retrieves the dependency of the requesting Bank Server and forces all of these Bank Servers to do checkpoint as well to have a consistent checkpoint.

## 4.2.2. Consistent Checkpoint

### 4.2.2.1. The purpose

When a Bank Server does a checkpoint, it requests to record it current state, which is stored in memory, to the physical storage, in this case is the Checkpoint Logging file. Since the time a new checkpoint is done, all operations occur are recorded into the Message Logging file. If Bank Server crashes, these operations can be replayed from the states saved in Checkpoint Logging file to resume perfectly the previous state. With the two files Checkpoint Logging file and Message Logging file, if crashed, each Bank Server can locally and independently resume its state without involve in any other enforcement.

Bank Server only requests to do checkpoint to record it new state and clean up the Message Logging file when this file has more that ten records.

As a plus, when a Bank Server requests to do a checkpoint, Recovery Module forces all the Bank Server dependent on the requesting Bank Server to do checkpoints as well. If so, checkpoint operation has to be synchronized among all these dependent Bank Server to guarantee checkpoints taken at consistent time at dependent Bank Servers.

### 4.2.2.2. The mechanism

If Bank Server does a transfer operation to Bank Server j, we say it is a dependency between the two banks. Since the new state of these two banks have not been saved yet, the leading Bank Server, Bank Server i, sends a message "DEPENDENCE BankServer#i-BankServer#j" to the Recovery Module. If there is an existing dependency containing either or both Bank Servers, this dependency will join the existing dependency.

The dependency is maintain so that when a Bank Server request to do a checkpoint to record it new state, all dependent Bank Servers records it new state as well.

Recovery Module use 2-phase commit to do a consistent checkpoint for all Bank Servers of a dependency. At phase one, Recovery Module sends message "READY_FOR_CHECKPOINT" to all the Bank Server in the current dependency. The Bank Servers send back message "READY_FOR _CHECKPOINT" if they are still alive. At phase two, Recovery Module sends "DO_CHECKPOINT". The Bank Servers do checkpoint operation and send back message "CHECKPOINT_DONE". This operation is explained in more details in later part.

In either phase one or phase two, if one of the Bank Server crashes, Recovery Module will not collect enough responses from all the Bank Servers. Recovery Module will cancel the checkpoint operation, send message "CANCEL_CHECKPOINT" to Bank Server who are already ready for checkpoint so that all these Bank Server stop waiting and go back to other works. This is the case for canceled checkpoint operation.

The communication between Recovery Module and Bank Servers uses TCP connection, which is a reliable connection. In most of the case, we assume the sending and receiving message achieve. For any reason that the communication fails, the time out occurs in both sending and receiving message. It will be treated as the canceled checkpoint operation.

The Recovery Module rarely crashes. But if it does, the time out in receiving message from the Bank Server occurs as a result for TCP connection. It also is treated as the canceled checkpoint operation.

So in any case, the checkpoint operation will achieve for all to gain a consistent checkpoints for all Bank Server in a dependency or cancel all.

### 4.2.3. Pool of dependency

A dependency is a vector containing any Bank Server having operations involved two sides, in this case is the transfer operation.

The pool of dependency is maintained by Recovery Module, which is a collection of all currently vectors of dependency. The dependency only is maintained and has meaning in the time from an old checkpoint and a new checkpoint is taken. This dependency is used to force all Bank Servers dependent on each other to do checkpoint to have a consistent clean cut. After the checkpoint is taken, the dependency is released.

### 4.2.3.1. Add a new dependency

When Bank Server i finished transfer operation with Bank Server j, the leading Bank Server i sends to Recovery Module a dependency message "DEPENDENCY BankServer#i-BankServer#j". Recovery Module looks at all the vector of dependencies:

- If both Bank Server i and Bank Server j has not found in any dependencies in the pool, Recovery Module create a new dependency with BankServer#i BankServer#j.
- If Bank Server i is found in a dependency but Bank Server j is not found in any dependency, Recovery Module adds Bank Server j into that dependency.
- If Bank Server j is found in a dependency but Bank Server i is not found in any dependency, Recovery Module adds Bank Server i into that dependency.
- If Bank Server i is found in a dependency and Bank Server j is found in another dependency, Recovery Module joins the first and the second into one dependency.

The management allows all Bank Server dependent directly or indirectly on each other is maintain in one dependency. So when one of them request to do a checkpoint, all Bank Servers in this dependency to be forced to do checkpoint as well.

### 4.2.3.2. Remove a dependency

A dependency is only maintained for the purpose of doing a consistent checkpoint. When the consistent checkpoint is done successfully for all Bank Servers of a dependency, Recovery will automatically remove that dependency from the pool.

### 4.2.4. FIFO channel

The channel of communication needs to be FIFO to avoid the latter request of checkpoint will not overlap or finish before the first request.

As mention above, the Recovery Module has two ports. One is used to keep listening to general requests. If that request needs more communication for completion the task, the second port is used to this purpose. So the first port is fee up and back to listen to other request rather than be holding for continuing for the first request.

If two consecutive "CHECKPOINT" requests come to the first port, Recovery Module reads the first checkpoint request, then use the second port for 2-phase commit checkpoint. At this time, all dependent Bank Servers are freeze for doing the checkpoint.

Also the second port is busy with this job. Thus, the second checkpoint request is queue up in the first port.

It's more technique to explain, but the result is guarantee for the FIFO channel.

## 4.2.5. Message to Recovery Module

Recovery Module listens to all requests from Bank Servers and Monitor. There are three kinds of messages coming to Recovery Module. They are message "CHECKPOINT", message ""DEPENDENCY" or message "RESTART".

### 4.2.5.1. RM – BS Message

**Message "CHECKPOINT"**

**Usage:** When a Bank Server has more than ten records in its message logging file and wishes to clean up the message logging file, Bank Server sends message "CHECKPOINT" to Recovery Module.

Recovery Module uses two phase commit to do the checkpoint for all dependent Bank Servers.

**Format:** CHECKPOINT

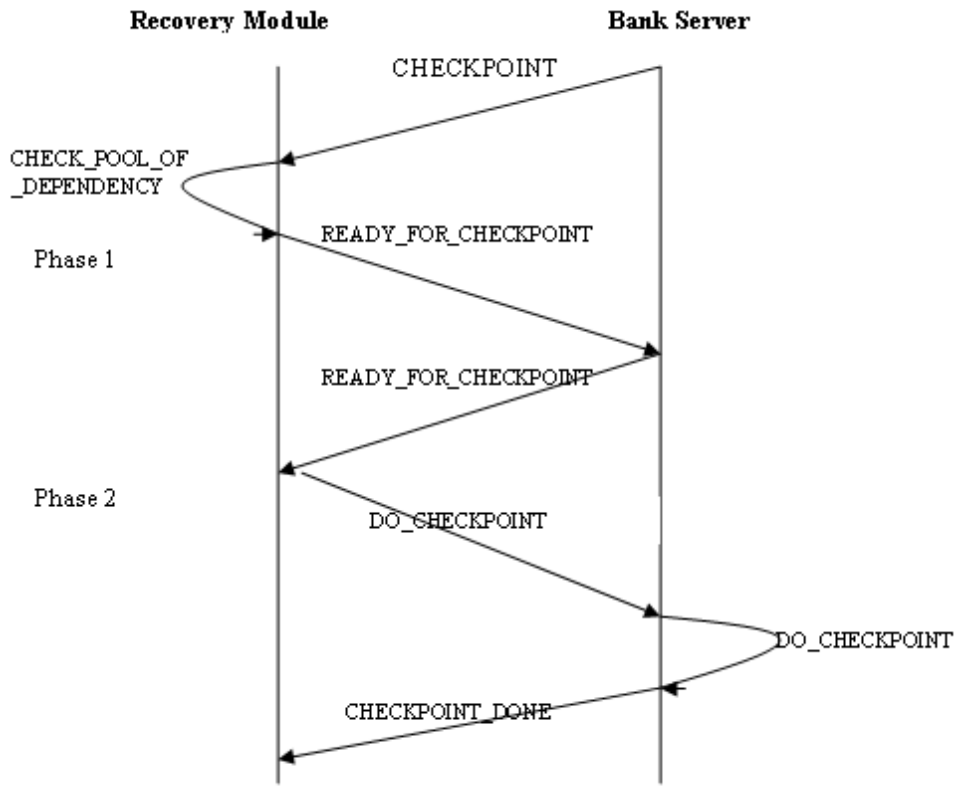

**Figure 6: 2-phase commit checkpoint with an independent Bank Server**

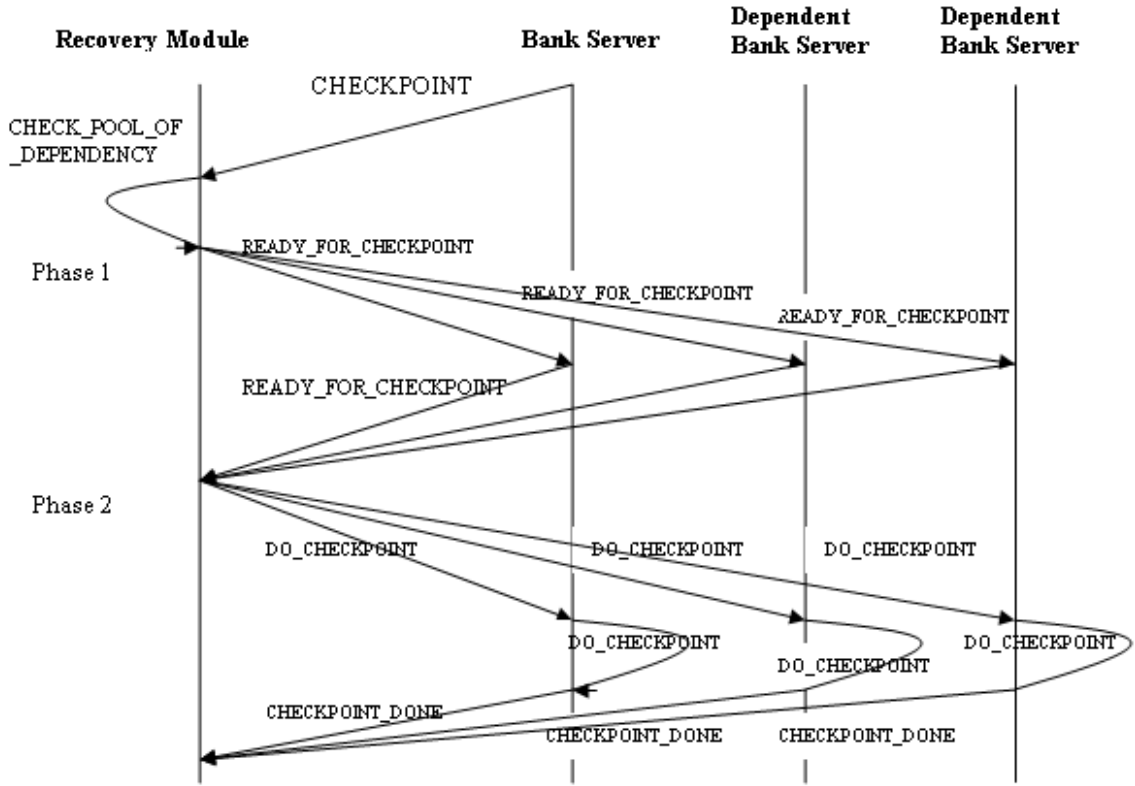

**Figure 7: 2-phase commit checkpoint with Bank Servers of a dependency**

**Message "DEPENDENCY"**

**Usage:** When the transfer operation between two Bank Servers is finished, the leading Bank Servers sends message "DEPENDENCY BankServer#i-BankServer#j" to Recovery Module. And Recovery Module manages this dependency as describe in the above part – pool of dependency.

**Format:** DEPENDENCY

### 4.2.5.2. RM – MM Message

**Message "RESTART"**

**Usage:** Monitor Module keeps monitoring all Bank Servers to check if they are still alive. If Monitor detects any Bank Server crashed, Monitor informs message "RESTART" to Recovery Module. The Recovery Module just simple restarts that Bank Server.

**Format:** RESTART

## 4.2.6. Design

### 4.2.6.1. Use case

#### 4.2.6.1.1. Use case Checkpoint

*Use case: Checkpoint*

Preconditions: Message Logging file of Bank Server i contains more than ten records

Postconditions: The checkpoint process is performed.

**Main Success Scenario:**
1. Bank Server i sends message "CHECKPOINT" to Recovery Module.
2. Recovery Module first checks if the requesting Bank Server belongs to the pool of dependency.
3. Recovery Module retrieves the vector of dependency containing the requesting Bank Server.
4. Recovery Module sends "READY_FOR_CHECKPOINT" to the Bank Server in the vector of dependency.
5. If the received Bank Server is ready, it sends message "READY_FOR_CHECKPOINT" back to Recovery Module.
   *Recovery Module repeats step 4-5 until it collects responds from all Bank Servers in the current vector of dependency.*
6. Recovery sends "DO_CHECKPOINT" to the Bank Server in the vector of dependency.
7. The received Bank Server performs <u>DoCheckpoint</u> user case
8. The received Bank Server sends message "CHECKPOINT_DONE" to Recovery Module
   *Recovery Module repeats step 6-8 until it collects responds from all Bank Servers in the current vector of dependency.*
9. Recovery yields to finish the checkpoint task.

**Extensions:**
3a. If there is no dependency for the requesting Bank Server.
      3.1. Recovery Module creates a new vector of dependency with the requesting Bank Server and stores it in the pool of dependencies.
      3.2. Recovery Module retrieves the recent created vector of dependency.
5a. Recovery Module don't receive responses from all dependent Bank Server – one of the Bank Server crash or the sending message did not reach the dependent Bank Server
    1. Recovery Module throws exception
    2. Recovery Module sends message to Bank Server who already said "READY_FOR_CHECKPOINT" that "CANCEL_CHECKPOINT" message
    3. The process has to stop. The Bank Server maintains the old checkpoint.
8a. Recovery Module don't receive responses from all dependent Bank Server – one of the Bank Server crash or the sending message did not reach the dependent Bank Server
    1. Recovery Module throws exception that checkpoint can not be done for all Bank Servers.

**4.2.6.1.2. Use case Dependency**
*Use case: Dependency*
Preconditions: Bank Server i successfully finishes transfer operation with Bank Server j
Postconditions: The pool of dependencies is update.
**Main Success Scenario:**
1. Bank Server sends message "DEPENDENT BankServer#i-BankServer#j" to Recovery Module
2. Recovery Module looks for this dependency in the pool of dependencies
3. Recovery Module creates a new dependency for BankServer#i-BankServer#j if each of them has not belonged to any other dependency.

4. Recovery Module adds this new dependency into the pool of dependencies
5. Print out the current dependency vector

**Extensions:**

3a. Bank Server i belonged to a dependency v1, but Bank Server j has not belonged to any dependency.

      a. Add BankServer#j to vector dependency v1

      b. Go to 5.

3b. Bank Server j belonged to a dependency v2, but Bank Server j has not belonged to any dependency.

      3.1. Add BankServer#i to vector dependency v2

      3.2. Go to 5.

3c. Bank Server i belonged to a vector dependency v1, Bank Server j has not belonged to vector dependency v2.

      3.1. Join vector dependency v1 and vector dependency v2.

      3.2. Go to 5.

### 4.2.6.1.3. Use case Restart

*Use case: Restart*

Preconditions: A Bank Server i crashed

Postconditions: The Bank Server crashed is restarted.

**Main success scenario:**

1. Monitor sends message "RESTART BankServer#i" to Recovery Module
2. Recovery Module restarts the Bank Server crashed

**Extension:**

2a. An error happened

      1. Recovery Module throws an error exception.

### 4.2.6.2. Class diagram

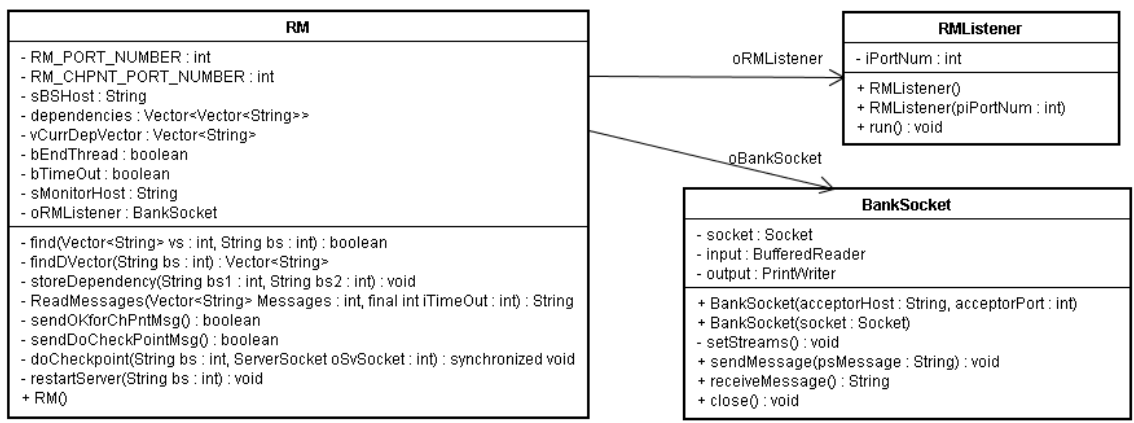

**Figure 8: Class Diagram of Recovery Module Component**

### Class BankSocket

This is a socket class for TCP connection. It creates a socket for sending and receiving messages from others banks.

**Class RMListener**
This class implements the Recover Module Listener as threads. Recovery Module listens requests from other Bank Servers and Monitor. Based on the received messages, Recovery Module delegates to the appropriate function to handle the message.
If the received message is "CHECKPOINT", Recovery Module calls function doCheckpoint to start the 2-phase commit checkpoint for the requesting Bank Server and other dependent Bank Server if applicable.
If the received message is "RESTART", Recovery Module calls function restartServer to restart the Bank Server crashed.
If the received message is "DEPENDENCY", Recovery Module calls function storeDependency to handle the dependency of this relationship.

**Function private boolean find(Vector<String> vs, String bs)**
This function looks for a specific Bank Server named *bs* in a dependency vector *vs* in the pool of dependency. The result returns *true* if bs is found in vs. Otherwise, the returned value is *false*.
The pool of dependency is maintained by the private variable *dependencies* of class RM. *Dependencies* is the vector containing all dependency vectors. Each dependency vector is a collection of Bank Server dependent on each other. Bank Server *bs* is search in each dependency vector *vs* to find if it belongs to any of the existing dependency.

This function is also used to look if a string bs is in a vector string vs. Function ReadMessages uses this function to see if the received message match with the expecting value.

**Function private Vector<String> findDVector(String bs)**
This function retrieves each vector in the dependencies collection. Then it calls for function find(Vector<String>, String) to look for the Bank Server named bs in that vector.

**Function private void storeDependency(String bs1, String bs2)**
This function manages the *dependencies* pool when a new dependency bs1-bs2 coming.
If bs1-bs2 has not belonged to any dependency vector, create a new dependency vector, add these two Bank Server bs1, bs2 into that vector and add this new vector into the dependencies collection.
If bs1 belonged to a dependency vector but bs2 has not belonged to any dependency vector, then add bs2 into the vector that bs belonged to and vice versa.
If bs1 belongs to a dependency vector and bs2 belongs to another, then join these two vectors.

**Function private String ReadMessages(Vector<String> Messages, final int iTimeOut) throws IOException**
This function is used to read message between the Recovery Module and a Bank Server at a private port 3001 of Recovery Module. The message is read at run time.

The received message is compared with the expecting value *Message*. If they match each other, then return the received message. Otherwise, a null value is return.

**Function private synchronized void doCheckpoint(String bs, ServerSocket oSvSocket)**
This function is used to start the 2-phase commit checkpoint when a Bank Server requests to do a checkpoint.
It finds the instance of the dependency vector containing all Bank Server dependent on the requesting Bank Server. This dependency vector is stored in *vCurrDepVector*, a private variable of class RM. If the requesting Bank Server is not dependent on any other Bank Server, create a new dependency for it. And the checkpoint is done for only this requesting Bank Server. Otherwise, retrieve the dependency vector and do the checkpoint for all these Bank Severs.
It starts the 2-phase protocol by calling function sendOKforChPntMsg for the first phase. If the first phase is fine, then it continues to call function sendDoCheckPointMsg for the second phase. If the second phase is fine, then it removes this dependency from the pool of dependencies. Anytime within these two phases, if a error happens or the 2-phase fails, the exception is thrown in these two functions.

**Function private boolean sendOKforChPntMsg()**
This function implements the first phase of the 2-phase commit protocol.
RM starts to send message "READY_FOR_CHECKPOINT" to the dependent Bank Server and waits for the responds from them. It's done sequently for all Bank Servers in the *vCurrDepVector* vector. The sending and reading message is done at run time.
If this process is done perfectly meaning RM receives responds from all Bank Server. Otherwise, if an error happens such as the sending and receiving fails or the Bank Server crashes, an exception is thrown that the checkpoint is canceled. In this case, RM sends message back to Bank Server who is ready for checkpoint to cancel the checkpoint.

**Function private boolean sendDoCheckPointMsg()**
This function implements the second phase of the 2-phase commit protocol.
RM starts to send message "DO_CHECKPOINT" to the dependent Bank Server and waits for the responds from them. It's done sequently for all Bank Servers in the *vCurrDepVector* vector. The sending and reading message is done at run time.
If RM receives responds from all Bank Server, the second phase is done successfully. Otherwise, if an error happens such as the sending and receiving fails or the Bank Server crashes, an exception is thrown that the checkpoint is canceled.

**Function private void restartServer(String bs)**
This function is used to restart a Bank Server crashed.
RM reads in the bs name passed by the Monitor. RM restarts that Bank Server.

**Function public RM()**
This function is a constructor for RM which initializes RM with the general port 3000 to listen to all requests.

## 4.3. Monitor

### 4.3.1. The Monitor Process

In this process, Monitor Module establishes a connection with Recovery Module as well as each bank server at first, and then keeps listening at corresponding ports for receiving heartbeat messages from existing active bank servers as well as register messages from new or restarting bank servers. If certain bank server crashes, Monitor Module sends a restart message to Recovery Module to restart the crashed bank server. The following sections will describe the monitor process in detail.

### 4.3.2. Message Mechanism

In our design, we use message mechanism to provide an effective monitoring service for the whole DBS, since all the DBS inner-module communication goes through messages. There are three kinds of messages within the Monitor Module: the first one is the heartbeat message for checking the status of existing active bank servers; the second one is the register message for registering new or restarting messages, and the third one is the restart message for informing Recovery Module to restart a crashed bank server.

Comparing to the other two modules in the DBS, the messages within Monitor Module are not be logged on a stable storage. Monitor Module processes those messages as soon as it receives them, and the socket communication mechanism will handle remaining concerns, such as message queue, garbage collection.

#### 4.3.2.1. Message "Heartbeat"

- This message takes place in the communication between Monitor Module and bank servers;
- It is sent by each bank server to the Monitor Module between a predefined time interval;
- The message notifies the Monitor Module that the sender is an working bank server and does not need to be restarted.
- If a bank server fails to send this message within the time interval, the Monitor Module will request the Recovery Module to restart that bank server, since the Monitor Module considers that bank server crashes and needs to be restarted.
- This message has the following format: HEARTBEAT_MSG {BS#}. The first segment indicates the type of message which is heartbeat message, and the second segment indicates the ID of bank server that sends this message.

#### 4.3.2.2. Message "Register"

- This message takes place in the communication between the Monitor Module and bank servers.
- It is sent by each bank server at its starting time in order to register with the Monitor Module as a sender of heartbeat messages.
- This message has the following format: REGISTER_MSG {BS IP address} {BS#}. The first segment indicates the type of message which is regiter message; the second segment indicates the IP address of bank server that sends this message, and the third segment indicates the ID of this bank server.

### 4.3.2.3. Message "Restart"

- This message takes place in the communication between the Monitor Module and Reconvery Module.
- It is sent by Monitor Module to Recovery Module when certain bank server crashes.
- This message has the following format: RESTART. The only segment indicates the type of message which is restart message.

## 4.3.3. Design

### 4.3.3.1. Use Case Diagram

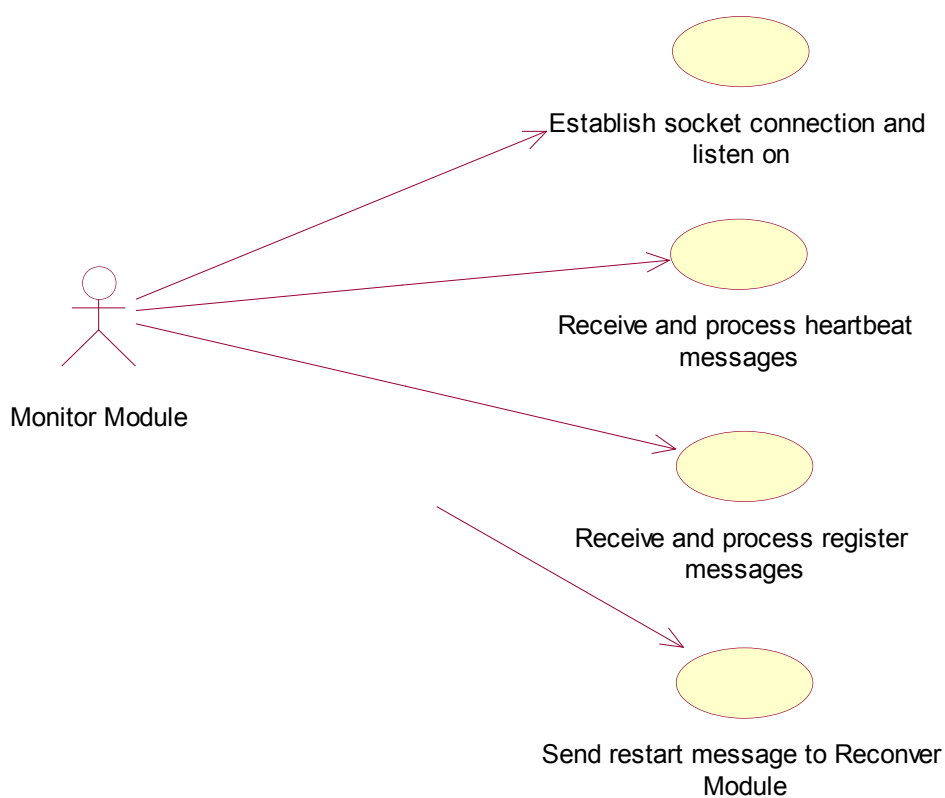

**Figure 9: Use Case Diagram of Monitor Module**

- Use Case "Establish socket connection and listen on": 1) Preconditions: a Monitor object is initiated; 2) Postconditions: the socket connection between Monitor object and each existing active bank server is established; the Monitor object begins to listen on heartbeat messages and register messages at each specified port;
- Use Case "Receive and process heartbeat message": 1) Preconditions: Monitor object receives heartbeat messages from existing active bank servers; 2) Postconditions: updates the corresponding records in the hash table of heartbeat messages according to their receiving time;

- Use Case "Receive and process register message": 1) Preconditions: Monitor object receives heartbeat messages from new or restart bank servers; 2) Postconditions: the socket connection between Monitor object and each new or restart bank server is established;
- Use Case "Send restart message to Recovery Module": 1) Preconditions: the time lapse of certain bank server since last heartbeat message exceeds the predefined time limit; 2) Postconditions: the Monitor object sends a restart message to Recovery Module to restart the crashed bank server.

### 4.3.3.2. Sequence Diagram

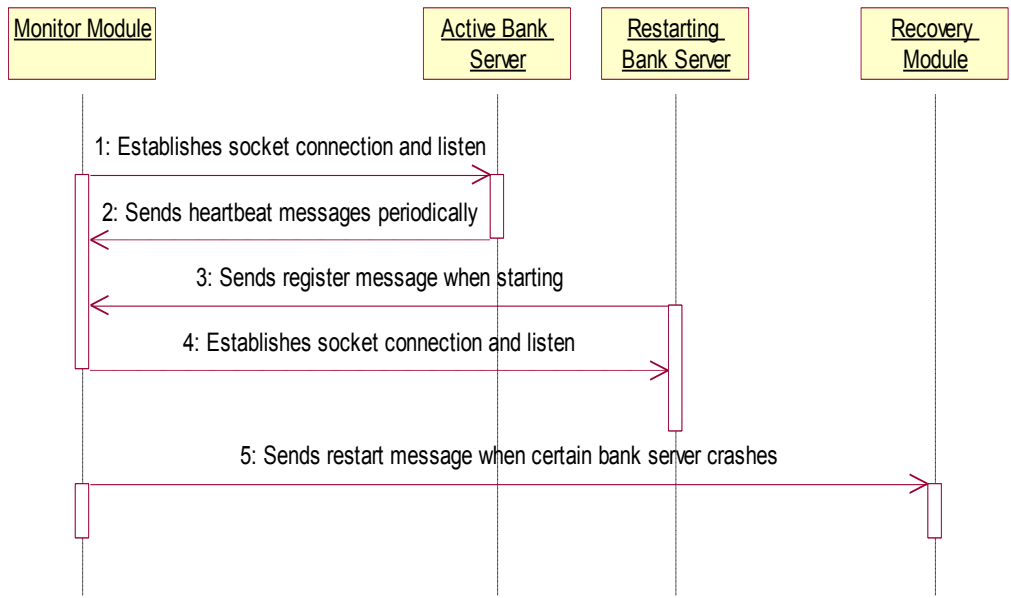

**Figure 10: Sequence Diagram of Monitor Module**

**4.3.3.3. Class Diagram**

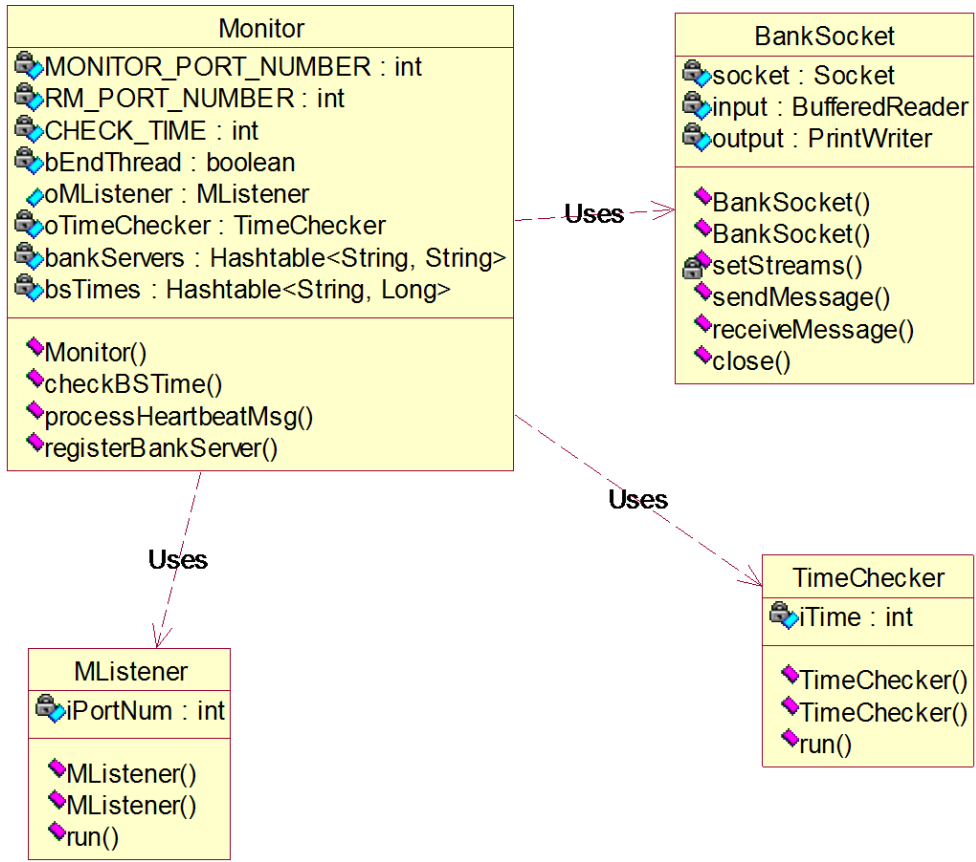

**Figure 11: Class Diagram of Monitor Module**

● Class "Monitor": This class implement the functionalities of Monitor Module;
  <u>Function private synchronized void checkBSTime()</u>: makes the decision whether sends a restart message to Recovery Module;
  <u>Function private synchronized void processHeartbeatMsg(String bsNumber)</u>: processes received heartbeat messages;
  <u>Function private synchronized void registerBankServer(String bsHost, String bsNumber)</u>: processes received register messages;
  <u>Function public Monitor():</u>the constructor for this class.

● Class "BankSocket": This class wraps the Java Socket class, by providing *sendMessage* and *receiveMessage* functions. The class internally maintains the input and output buffers and takes precautions for buffer flushing;

● Class "MListener": This class instantiates threads used for establishing socket connection with bank servers;
  <u>Function public void run()</u>: 1) instantiates a socket for accepting connection; 2) runs a loop for reading incoming heartbeat and register messages;
  <u>Function MListener ()</u>: the constructor for this class.

● Class "TimeChecker": This class instantiates threads that check the heartbeat message time for every registered bank server;
Function public void run(): checks the heartbeat message time whether exceed predefined time limit for each bank server;
Function TimeChecker(): the constructor for this class.

# 5. Conclusion

By consolidating the Bank Server and building up the middle ware with Recovery Module and Monitor Module, we add the feature of fault tolerance into the BDS system, which allows the system automatically to recover from failures. Moreover, after restarted by Recovery Module, and with the message logging and checkpoint logging protocol, Bank Server can independently and locally resume its previous state and be able to process subsequent requests for the various operations.